\documentclass[sigconf]{acmart}

\usepackage{graphicx}
\usepackage{bm}
\usepackage{subfig} 
\usepackage{hyperref}
\usepackage{amsmath}
\usepackage{multirow}
\usepackage{xcolor}
\usepackage{comment}
\usepackage{enumitem}
\usepackage{soul}
\usepackage{colortbl}
\usepackage{natbib}
\usepackage{tikz}
\usepackage{lipsum}

\newlength{\RoundedBoxWidth}
\newsavebox{\GrayRoundedBox}
\newenvironment{GrayBox}[1][\dimexpr\linewidth-4.5ex]%
   {\setlength{\RoundedBoxWidth}{\dimexpr#1}
    \begin{lrbox}{\GrayRoundedBox}
       \begin{minipage}{\RoundedBoxWidth}}%
   {   \end{minipage}
    \end{lrbox}
    \begin{center}
    \begin{tikzpicture}%
       \draw node[draw=black,fill=black!10,rounded corners,%
             inner sep=2ex,text width=\RoundedBoxWidth]%
             {\usebox{\GrayRoundedBox}};
    \end{tikzpicture}
    \end{center}}

\copyrightyear{2022} 
\acmYear{2022} 
\setcopyright{acmcopyright}\acmConference[ARES 2022]{The 17th International Conference on Availability, Reliability and Security}{August 23--26, 2022}{Vienna, Austria}
\acmBooktitle{The 17th International Conference on Availability, Reliability and Security (ARES 2022), August 23--26, 2022, Vienna, Austria}
\acmPrice{15.00}
\acmDOI{10.1145/3538969.3538986}
\acmISBN{978-1-4503-9670-7/22/08}

\begin{document}

\title[SoK: Security of Microservice Applications]{SoK: Security of Microservice Applications: A Practitioners' Perspective on Challenges and Best Practices} 

\author{Priyanka Billawa}
\affiliation{%
  \institution{Hamburg University of Technology} 
  \city{Hamburg} \country{Germany}
  }
\email{priyanka.billawa@tuhh.de}

\author{Anusha Bambhore Tukaram}
\affiliation{%
  \institution{Hamburg University of Technology} 
  \city{Hamburg} \country{Germany}
  }
\email{anusha.bambhoretukaram@tuhh.de}

\author{Nicolas E. Diaz Ferreyra}
\affiliation{%
  \institution{Hamburg University of Technology} 
  \city{Hamburg} \country{Germany}
  }
\email{nicolas.diaz-ferreyra@tuhh.de}

\author{Jan-Philipp Stegh\"{o}fer}
\affiliation{%
  \institution{Chalmers University of Technology $|$ University of Gothenburg} 
  \city{Gothenburg} \country{Sweden}
  }
\email{jan-philipp.steghofer@cse.gu.se}

\author{Riccardo Scandariato}
\affiliation{%
  \institution{Hamburg University of Technology} 
  \city{Hamburg} \country{Germany}
  }
\email{riccardo.scandariato@tuhh.de}

\author{Georg Simhandl}
\affiliation{%
  \institution{University of Vienna} 
  \city{Vienna} \country{Austria}
  }
\email{georg.simhandl@univie.ac.at}

\renewcommand{\shortauthors}{Billawa et al.}

\begin{abstract} 
  Cloud-based application deployment is becoming increasingly popular among businesses, thanks to the emergence of microservices. However, securing such architectures is a challenging task since traditional security concepts cannot be directly applied to microservice architectures due to their distributed nature. The situation is exacerbated by the scattered nature of guidelines and best practices advocated by practitioners and organizations in this field. In this research paper we aim to shay light over the current microservice security discussions hidden within Grey Literature (GL) sources. Particularly, we identify the challenges that arise when securing microservice architectures, as well as solutions recommended by practitioners to address these issues. For this, we conducted a systematic GL study on the challenges and best practices of microservice security present in the Internet with the goal of capturing relevant discussions in blogs, white papers, and standards. We collected 312 GL sources from which 57 were rigorously classified and analyzed. This analysis on the one hand validated past academic literature studies in the area of microservice security, but it also identified improvements to existing methodologies pointing towards future research directions.
\end{abstract}



\maketitle

\section{Introduction} \label{intro}

Microservices can be seen as small autonomous services that interact with each other to achieve a specific business goal. They are characterized by a high abstraction level and decentralization with the purpose of isolating failures as well as making services independently deployable and highly observable \cite{newman2021building}. Several benefits are linked to microservice-based architectures such as increased scalability, flexibility, and agility. However, breaking down a system into individual, independent units increases the complexity of securing it by folds \cite{yarygina2018overcoming}.

Security is often viewed as an afterthought in the implementation of software projects. This has proven to be inefficient even when developing traditional monolithic architectures. When addressing microservice architectures, considering security late in development is prohibitive~\cite{yarygina2018overcoming}. This is in part due to their widely distributed interfaces that expand the attack surface of a system by exposing more data and information about its endpoints \cite{ahmadvand2016requirements}. Moreover, as monolithic applications services are migrated into microservices, code that was not supposed to be externally accessible is now exposed through web APIs~\cite{yarygina2018overcoming}. Therefore, there is a need for frameworks and best practices that take security into consideration from the early stages of microservices' life cycle.

\subsection*{Motivation}

Many security efforts from academia have been documented within the white literature (c.f., \cite{yarygina2018overcoming,pereira2019security,hannousse2021securing}). The software industry has also contributed actively to ongoing research on microservice security, often disseminated as \textit{grey literature}~(GL), e.g., as reports, proceedings, and articles that are not formally published nor available in a commercial way. Nevertheless, resources published by industrial practitioners do not have as much dissemination and visibility as academic publications \cite{paez2017gray}. Acknowledging GL sources can (i) help increase the awareness and comprehensiveness of the microservice research landscape, and (ii) foster a more balanced and unbiased body of knowledge in this regard. Therefore, examining GL can help us identify the shortcomings in our current understanding of microservice security (which is mostly based on academic research) as well as discovering new ideas and methodologies proposed by software practitioners. 

\subsection*{Contribution and Research Questions}

The goal of this work is to gain insight into the security challenges and countermeasures frequently reported by microservices practitioners. Particularly, it aims at (i) documenting the \textit{challenges} encountered in securing microservice architectures, and (ii) identifying \textit{solutions} implemented by microservice developers to address such challenges in practice. For this, we conducted a GL review of 312 sources from which 57 were closely and rigorously analyzed.

To establish a common understanding of the terms used in our study, we defined \textit{challenges} as undesirable situations encountered during development that may reduce a system's ability to deal with attacks, thus preventing it from achieving its security goals. \textit{Solutions} on the other hand were defined as ideas and workarounds identified by practitioners to overcome these challenges. Using this knowledge, we provide a big picture of the current state of microservice-based architectures security. Therefore, we elaborate on the following research questions:
\begin{itemize}[itemsep=3pt,topsep=0pt,leftmargin=*]
    \item[\textbullet] \textit{\textbf{RQ1}: What are the current challenges reported by practitioners in the field of microservice security?} \\[3pt]
    The purpose of this research question is to identify potential challenges and issues addressed by practitioners working on microservice-based architectures. This also includes non-technical stakeholders (such as managers, product owners and entrepreneurs) who may be responsible for key decisions when discussing project requirements. 
    \item[\textbullet] \textit{\textbf{RQ2}: How do practitioners address the challenges mentioned in RQ1 and what are their recommendations to overcome these challenges?}\\[3pt] This research question seeks to understand the various solutions proposed by the practitioners to overcome above challenges. As these solutions can be either generic or technical, we further refine RQ2 as:
    \begin{itemize}[itemsep=3pt,topsep=3pt]
       \item[\textendash] \textit{\textbf{RQ2a}: What best practices are mentioned by practitioners?}\\[3pt]
       Here we look for generic advice in the form of best practices recommended by the industry that need to be followed when dealing with microservice architectures. For example, in order to avoid man-in-the-middle attacks, secrets must not be shared over a communication channel in unencrypted form. 
       \item[\textendash] \textit{\textbf{RQ2b}: What technical solutions do practitioners propose?}\\[3pt]
       Here we seek to identify the technical solutions (improvements/workarounds) implemented by practitioners to overcome challenges. Technical solutions take the form of libraries, code, standards, tools and patterns. For instance, OAuth 2.0 is a standard recommended frequently for authorization.
    \end{itemize}
\end{itemize}

The results of this study provide an overview of the state of practice of microservice security in the field. Particularly, it helps acknowledging the security challenges practitioners often come across when developing microservice-based architectures, as well as the best practices they follow to overcome them. The outcome of this work should also serve as a reference guide for academics and software practitioners seeking to envisage and shape new research paths and directions.

The rest of this paper is organized as follows. Section~\ref{sec:related-work} gives of an overview of the related work in white and GL in microservice security. In Section~\ref{sec:methodology}, we present the research methodology followed by a discussion of the results of the study in Section~\ref{sec:results}. Section~\ref{sec:overlooked_areas} elaborates on overlooked areas in microservice security research, whereas Section~\ref{sec:threats-to-validity} is concerned with the limitations of this study. Finally, the conclusion is presented in Section~\ref{sec:conclusion}.

\section{Related Work}
\label{sec:related-work}
In 2018, Soldani et al.~\cite{soldani2018pains} conducted the first systematic mapping of GL on securing microservices. This study focused on the \textit{pains} and \textit{gains} of the microservice architectural style. Here security was presented as a design concern which provides \textit{gains} in terms of automation, fine-grained policies, fire walling, isolation, and layering. Conversely, it was also seen as a trigger of \textit{pains} like access control, centralised support, CI/CD, endpoint proliferation, human errors, and size/complexity. Nevertheless, this work was primarily focused on microservice Application Programming Interfaces (APIs). 

Another systematic literature review was conducted by  
Pereira et al.~\cite{pereira2019security} over a set of 26 academic sources published between 2015 and 2018. The study identified 18 security mechanisms proposed for microservice architectures and revealed that authentication and authorization were the most popular ones. These results were validated thereafter by comparing them against a set of security patterns found in a collection of open-source projects (c.f., \cite{marquez2018actual}). In line with this, Hannousse et al.~\cite{hannousse2021securing} contributed to the body of knowledge on secure microservices by analyzing 46 academic sources published between 2011 and 2020. The study resulted in an ontology of current microservice security threats and mechanisms. 

Still, a large portion of the body of knowledge in microservice security seems to be grounded in a small amount of academic publications, whereas the experiences gathered by industrial practitioners remain generally unknown. Hence, there is a call for acknowledging the technical contributions and findings present within the GL to expand our current understanding of security challenges and practices in microservice development.

Recently, work conducted by Ponce et al.~\cite{ponce2021smells} aimed to shed some light on the security trends in microservice deployment. The study is a multi-vocal literature review seeking to taxonomize microservice code smells and ways to refactor them. Although the study provides a useful list of code smells and ways to mitigate them, it does not address the challenges encountered in securing microservice architectures. Conversely, our study aims to complement the one by Ponce et al.~\cite{ponce2021smells} while answering further research questions concerning the tools and methods used in the practice.

\section{Research Methodology}
\label{sec:methodology}

We conducted a systematic review of GL sources following the guidelines proposed by Petersen et al.~\cite{petersen2015guidelines}. The different stages of this process are described in detail in the following subsections.

\begin{figure}
    \centering
    \includegraphics[width=0.9\linewidth]{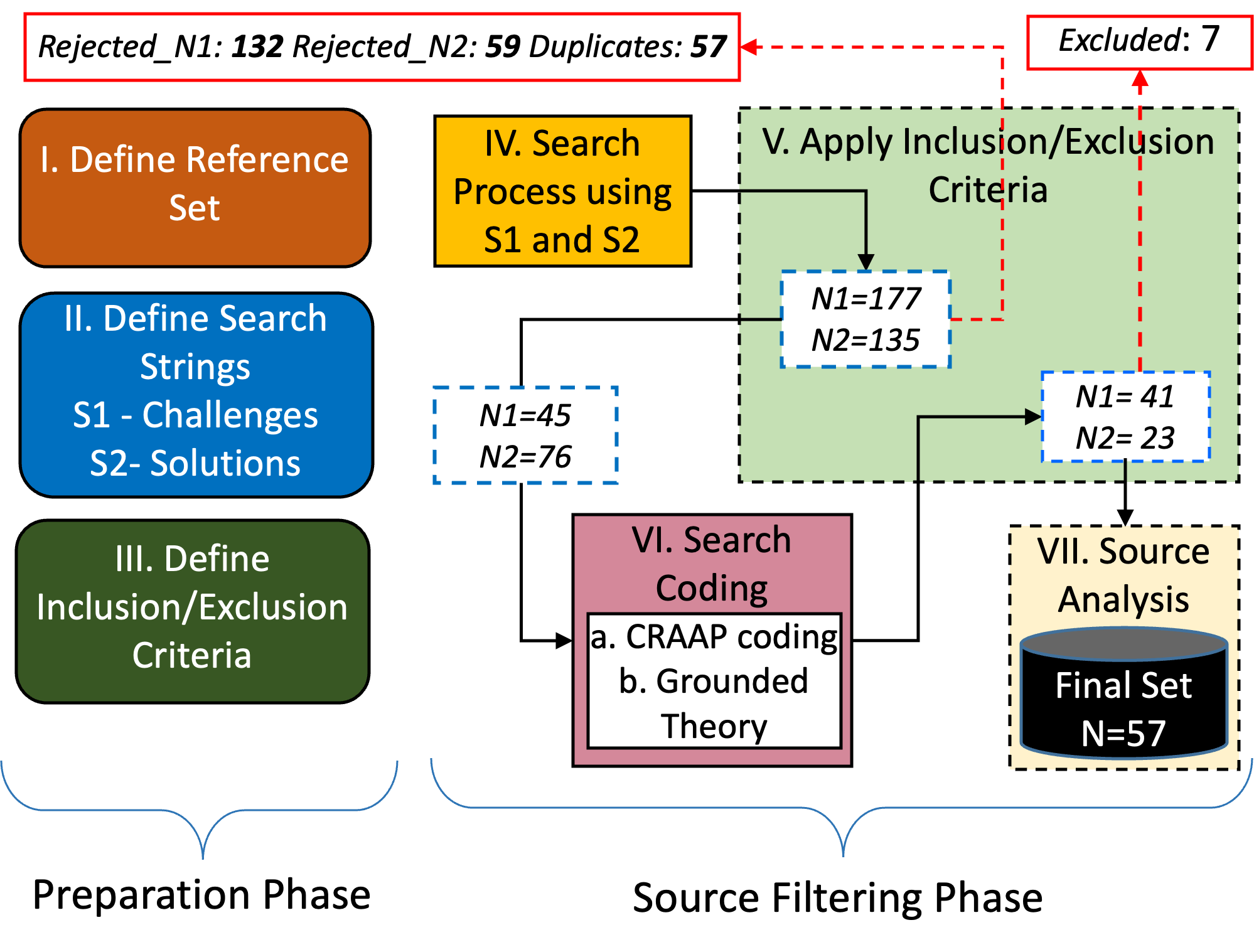}
    \caption{Research methodology overview. N1 and N2 correspond to the number of sources retrieved through S1 and S2 at each method step, respectively.}
    \label{workflow}
\end{figure}

\subsection{Preparation Phase} \label{preparation_phase}

As shown in Fig.~\ref{workflow}, the process starts with a preparation phase consisting of three sub-tasks: 

\subsubsection*{(i) Reference Set Definition:}

We created a reference set of 10 sources to setup a benchmark for the systematic retrieval of GL. It was compiled by performing an opportunistic search with simple terms related to microservice security (e.g., \texttt{microservice security solutions challenges}). It contains from 6 types of GL sources including \textit{standards}, \textit{blogs}, and \textit{coding community sites}. The complete reference set can be found in Table~S1 in the Supplementary Material\footnote{\url{https://tinyurl.com/greylitstudy}}.

\subsubsection*{(ii) Search String Formulation:}

The search strings used to retrieve the literature sources for this study were built out of a set of synonyms corresponding to the following categories:
\begin{itemize}
    \item \texttt{challenges}: \textit{code smells}, \textit{pitfalls}, \textit{vulnerability$^*$}, \textit{concern$^*$}, \textit{problem}, \textit{mistake}, \textit{trouble}, \textit{flaw$^*$}, \textit{issue$^*$}.
    \item \texttt{solutions}: \textit{best practice$^*$}, \textit{principles}, \textit{recommendation}, \textit{policy$^*$}, \textit{pattern$^*$}, \textit{guideline$^*$}, \textit{approach}, \textit{technique}, \textit{consideration}, \textit{strategy}.
\end{itemize}
On discussing the various possible combinations of relevant keywords from experience, we discarded some of the synonyms (only those marked with an `*' were further considered) and formulated a \textit{base string} \bm{$V_0$}: (\texttt{microservice} AND \texttt{security}) AND (\texttt{challenge} OR \texttt{pattern} OR \texttt{flaw} OR \texttt{practice}). Alongside, we derived four \textit{additional search strings} from $V_0$ by replacing \texttt{flaw} with \texttt{vulnerability} ($V_1$), \texttt{practice} with \texttt{guideline} ($V_2$), \texttt{practice} with \texttt{policy} ($V_3$), and \texttt{pattern} with \texttt{solution} ($V_4$). 

We used the reference set to assess whether the retrieved sources were of the expected type. Besides we compared the hits retrieved with each of the search strings $V_1$, $V_2$, $V_3$, and $V_4$ against $V_0$. We observed that the results gathered with \bm{$V_2$} contain the ones of $V_1$ and $V_3$ plus two more sources. On the other hand, \bm{$V_4$} generated additional sources when compared to $V_0$. To further examine the suitability of the search strings we conducted a follow-up analysis by combining $V_2$+$V_4$ into \bm{$V_5$}, and $V_0$+$V_2$+$V_4$ into \bm{$V_6$}. However, we observed very little overlap between the results generated by $V_5$, $V_6$, and $V_0$.

Based on the outcomes produced by each search string (and the said combinations) we concluded that (i) the presence of more keywords does not necessarily produce more precise results, and (ii) that the search engine (Google in this case) does not always generate the same outcome for a given set of synonyms. Hence, we decided to split the search query into a string containing keywords related only to \textit{challenges} ($S_1$), and another one only with keywords related to \textit{solutions} ($S_2$). Thereby we sought to generate a fairly large and diverse corpus of GL sources for further examination and processing.\\[0.5ex]

\fbox{\footnotesize \parbox{0.9\linewidth}{
\textbf{S1}:  (\texttt{microservice} AND \texttt{security}) AND (\texttt{challenge} OR \texttt{flaw} OR \texttt{concern} OR \texttt{issue} OR \texttt{vulnerability})
\\[2ex]
\textbf{S2}: (\texttt{microservice} AND \texttt{security}) AND (\texttt{pattern} OR \texttt{solution} OR \texttt{guideline} OR \texttt{practice} OR \texttt{policy}) 	
 }}

\subsubsection*{(iii) Defining the Inclusion/Exclusion Criteria:}

In order to skip irrelevant sources, we outlined a set of strict inclusion and exclusion criteria. Only sources meeting all the inclusion criterion (I) were considered for further examination. However, if a source met one of the exclusion criterion (E) it was automatically discarded. \autoref{inc} summarizes the inclusion and exclusion criteria used in our study, as well as the rationale behind them.

\begin{table}[]
\fontsize{12}{12}\selectfont
\centering
\setlength{\tabcolsep}{5pt}
\def\arraystretch{1.5}
\resizebox{\linewidth}{!}{%
\begin{tabular}{| m{2cm} m{8cm} | m{6cm} |}
\hline
\rowcolor[HTML]{D5D3D2} 
\multicolumn{2}{|c|}{\textbf{Inclusion Criteria}} &
  \multicolumn{1}{c|}{\textbf{Rationale}} \tabularnewline \hline
\multicolumn{1}{|l|}{I1} &
  Source provides information on microservice security challenges and means to overcome them &
  Collect evidence to support research questions \tabularnewline \hline
\multicolumn{1}{|l|}{I2} &
  Source provides unrestricted access (No fee) &
  Anyone interested can access the study \tabularnewline \hline
\multicolumn{1}{|l|}{I3} &
  The main topic of the source is Microservice Security &
  Collect evidence related to main focus of study \tabularnewline \hline
\rowcolor[HTML]{D5D3D2} 
\multicolumn{2}{|c|}{\textbf{Exclusion Criteria}} &
  \multicolumn{1}{c|}{\textbf{Rationale}} \tabularnewline \hline
\multicolumn{1}{|l|}{E1} &
  Source is a book, academic journal, document from an academic conference or patent &
  Beyond the scope of a GL review \tabularnewline \hline
\multicolumn{1}{|l|}{E2} &
  Source is a product brochure, advertisement (including call for papers) or product manual &
  Prone to bias and cannot be relied upon \tabularnewline \hline
\multicolumn{1}{|l|}{E3} &
  Source is inaccessible, unavailable or is a malicious link &
  Cf.~I2\tabularnewline \hline
\multicolumn{1}{|l|}{E4} &
  Source is an interview or bulletin reporting proceedings of academic conference &
  Information insufficient to answer research questions \tabularnewline \hline
\multicolumn{1}{|l|}{E5} &
  Source is too short for analysis (less that 2 minute read time) &
  Information insufficient to answer research questions \tabularnewline \hline
\multicolumn{1}{|l|}{E6} &
  Source does not mention Microservice security &
  Cf.~I3 \tabularnewline \hline
\multicolumn{1}{|l|}{E7} &
  Source is a tutorial, job posting, video, wiki, interview resource or LinkedIn profile &
  Serves a different purpose; does not help answer research questions \tabularnewline \hline
\multicolumn{1}{|l|}{E8} &
  Source content is a duplicate of another website &
  Does not contribute new insights \tabularnewline \hline
\end{tabular}%
}
\caption{Inclusion and exclusion criteria.}
\label{inc}
\end{table}

\subsection{Source Filtering Phase} \label{filtering_phase}

After having defined the search strings and the inclusion/exclusion criterion, we proceed with the search and filtering of GL sources:

\subsubsection*{(iv) Search process:}

The search was carried out according to the guidelines proposed by Garousi et al.~\cite{garousi2016need}. Here, the Google search engine is preferred due to its ability to retrieve a large number of relevant results. To ensure consistency, the search was conducted through a VPN tunnel provided by the \textit{**Anonymized University**} so that all searches originated at the same location. We also cleared all cookies and restarted the browser before running the search. We set the beginning of the search period to January 2011 as it corresponds to the year in which the term ``microservice architecture'' was coined\footnote{\url{https://martinfowler.com/articles/microservices.html}} and the search end was set to March 2021. The search stopped on saturation meaning that we included as many sources as possible until no more relevant results appeared (we judged the source relevance based on its title). By the end of this task a total of \textbf{312 hits} were retrieved from which 177 corresponded to search string S1 and 135 to S2.


\subsubsection*{(v) Apply Inclusion/Exclusion Criteria:} 

As shown in \autoref{workflow}, a total of 191 sources were discarded after applying the inclusion/exclusion criterion: 132 correspond to results generated by S1 and 59 by S2. Another 57 duplicated sources were skipped resulting in a corpus of \textbf{64 sources} that went through a further full-text assessment.

\subsubsection*{(vi) Search coding:}

One critical aspect of our analysis is the credibility of the sources being processed. Hence, we conducted a CRAAP\footnote{\url{https://researchguides.ben.edu/source-evaluation}} test on our corpus for assessing its overall reliability. Such a test consists of formulating questions around a set of quality attributes, namely \textit{Currency}, \textit{Relevance}, \textit{Accuracy}, \textit{Authority} and \textit{Purpose}. For instance, we checked for information biases and spelling mistakes when assessing the \textit{accuracy} of a source. We also checked the biography/profile of the author (when available) to determine their level of \textit{authority}. 
This process was performed by generating lists of pre-defined code-words for each of the CRAAP categories. For example, ``overview'', ``in depth'', and ``inadequate'', were used to code the \textit{relevance} (adequacy of explanation) of GL sources. A complete list of the CRAAP aspects under analysis and the corresponding code-words can be found in the supplementary material (Table~S3).

We performed an additional coding of each GL source following a ground theory approach with the purpose of capturing relevant information for answering the RQs. For this, we created an initial list of code words related to RQ1 (challenges) and RQ2 (solutions). Any new terms discovered during the coding process were added to the initial list. Likewise, discrepancies in the coded data such as non-uniform naming or two code-words referring to the same concept were manually removed. The results of the coding analysis were documented in a spreadsheet available in an online repository\footnote{\url{https://tinyurl.com/greylitstudy}}. Also, a \textit{criteria catalogue} containing the interpretation and meaning of code words is maintained to eliminate ambiguities and inconsistencies. Such a catalogue is also included in the said spreadsheet.

\subsubsection*{(vii) Source analysis:}

A full-text analysis conducted just after the coding step revealed that 7 sources had to be further excluded (i.e., based on the exclusion/inclusion criteria). As a result, \textbf{57 GL sources} were included in the final set. We assigned them labels from \textbf{S1 to S57} to facilitate their reference throughout the rest of the paper (\autoref{overview}).

Using the information extracted during the ground theory coding we proceeded further to answer the RQs. Furthermore, the contents of the spreadsheet were saved as an SQL database, which is accessible from the same GitLab location. SQL queries were formulated to extract the required information from the database.

\renewcommand{\citenumfont}[1]{S#1}
\renewcommand{\bibnumfmt}[1]{[S#1]}

\begin{table*}

\centering
\caption{Overview of primary studies selected for coding and analysis.}
\label{overview}
\resizebox{\linewidth}{!}{%
\begin{tabular}{|c|c|c|c|c|} 
\hline
\rowcolor[HTML]{D5D3D2} \textbf{Year} & \textbf{Title} & \textbf{Website} & \textbf{Type} & \textbf{Cite} \\ 
\hline
2019 & Microservices Security: Challenges and Best Practice & Neuralegion & Blog & [S1] \\ 
\hline
2019 & 10 Ways Microservices Create New SecurityChallenges & Kong & Blog & [S2] \\ 
\hline
2016 & Security Challenges in Microservice Implementations & Container Solutions & Blog & [S3] \\ 
\hline
- & Security In AMicroserviceWorld & OWASP & Presentation & [S4] \\ 
\hline
2020 & How to Secure Microservices Architecture & Security Intelligence & Blog & [S5] \\ 
\hline
2020 & How to Exploit a Microservice Architecture & Crashtest Security & Blog & [S6] \\ 
\hline
2016 & Security challenges presented by microservices & Techgenix & Article & [S7] \\ 
\hline
2020 & 4 Fundamental Microservices Best Practices & TechTarget & Article & [S8] \\ 
\hline
2018 & Security Approaches for Microservice Architectures & Open Networking Summit & Presentation & [S9] \\ 
\hline
2018 & Microservices Security: Big Vulnerabilities Come in Small Packages & Dzone & Article & [S10] \\ 
\hline
2020 & 13 Best Practices to Secure Microservices & Geekflare & Blog & [S11] \\ 
\hline
-  & 8 best practices for microservices app-sec & TechBeacon & Article & [S12] \\ 
\hline
2019 & Microservices Security & WhiteHat Security & Blog & [S13] \\ 
\hline
2020 & Security Patterns for Microservice Architectures & Okta Developers & Blog & [S14] \\ 
\hline
2019 & Microservices introduce hidden security complexity, analyst warns & ComputerWeekly.com & Article & [S15] \\ 
\hline
2018 & Microservices Security: Probably Not What You Think It Is & TheNewStack & Article & [S16] \\ 
\hline
2019 & How Kubernetes Vulnerabilities is About Securing Microservices & TheFabricNet & Blog & [S17] \\ 
\hline
2018 & Why securing containers and microservices is a challenge & CSO & Article & [S18] \\ 
\hline
- & API and Microservice Security & PortSwigger & Blog & [S19] \\ 
\hline
2020 & Microservices: Security and architectural issue for internal services & Stack Overflow & Q$\&$A Forum & [S20] \\ 
\hline
2019 & Security Strategies for Microservices-based Application Systems & NIST & Document & [S21] \\ 
\hline
2021 & Tips for minimizing security risks inyour microservices & AT $\&$T Business & Article & [S22] \\ 
\hline
2020 & Contrast-Security-OSS java-microservice-sample-apps & GitHub & Code Repository & [S23] \\ 
\hline
2020 & 7 best practices for microservices security & Sqreen & Blog & [S24] \\ 
\hline
2019 & The top 3 considerations when securing your microservices architecture & Citrix & Article & [S25] \\ 
\hline
- & 7 Best Practices of Microservices Security & Newizze & Blog & [S26] \\ 
\hline
2019 & Secure your Microservices & RedHat OpenShift & Blog & [S27] \\ 
\hline
2018 & Cyber Tech: How-To Secure Microservices and Containers & SecurityInfowatch & Article & [S28] \\ 
\hline
2020 & Microservices and Container Security: 7 Best Practices for 2020 & Portshift & Blog & [S29] \\ 
\hline
2020 & Microservice Architecture and its 10 Most Important Design Patterns & Medium & Blog & [S30] \\ 
\hline
2018 & Microservice and Container Security: 10 Best Practices & Apriorit  & Blog & [S31] \\ 
\hline
2019 & Patterns in Microservices Authentication with Client Certificates & OpenLogic & Blog & [S32] \\ 
\hline
2019 & Orchestration: Avoiding Container Vulnerabilities & ContainerJournal & Article & [S33] \\ 
\hline
2019 & Securing modern API- and microservices-based apps by design & Kobalt.io & Article & [S34] \\ 
\hline
2017 & Securing Microservices: The API gateway, authentication and authorization & Software Development Times & Article & [S35] \\ 
\hline
2015 & Microservices: Simple servers, complex security & InfoWorld & Article & [S36] \\ 
\hline
2018 & What Makes Microservices Secure? Learning From Successes and Failures & TestFort & Blog & [S37] \\ 
\hline
2016 & Microservices Architecture \textbar{} Rethinking Application Security & AVI Networks & Blog & [S38] \\ 
\hline
2016 & Securing Microservices (Part I) & Medium & Blog & [S39] \\ 
\hline
2017 & Securing REST microservices with Spring Security & Stack Overflow & Q\&A & [S40] \\ 
\hline
2019 & Best Practices for Implementing a Secure Application Container Architecture & Cloud Security Alliance & Document &        [S41] \\ 
\hline
2019 & Istio Security: Running Microservices on Zero-Trust Networks & StackRox & Blog & [S42] \\ 
\hline
- & How can an API endpoint identify and authenticate the client making a request? & Microservice API Patterns & Blog &         [S43] \\ 
\hline
2020 & Securing Modern API and Microservices-Based Apps by Design – Part 2 & Forward Security & Article &         [S44] \\ 
\hline
2019 & The Phantom Token Approach & Curity.io & Blog & [S45] \\ 
\hline
2018 & Securing microservice environments in a hostile world & NetworkWorld & Article & [S46] \\ 
\hline
2021 & Build Resilient, Secure Microservices with Microsegmentation & TheNewStack & Blog & [S47] \\ 
\hline
2017 & Security Standard - Microservices Architecture (SS-028) & DWP Standards & Document & [S48] \\ 
\hline
2018 & The Role of IAM in Microservices & SlideShare -WSO2 Inc & Presentation & [S49] \\ 
\hline
2020 & How to perform authorization + services also need auth checks individually? & Stack Exchange & Q\&A & [S50] \\ 
\hline
2017 & Microservices security with Oauth2 & Piotr's TechBlog & Blog & [S51] \\ 
\hline
- & Easily Secure your Microservices with Keycloak & RedHat & Presentation & [S52] \\ 
\hline
2020 & Microservices Authorization using Open Policy Agent and Traefik & AppSecCo-Medium & Blog & [S53] \\ 
\hline
2018 & How a Service Mesh Can Help With Microservices Security & christianposta.com & Blog & [S54] \\ 
\hline
2018 & Microservices Authentication and Authorization Best Practice & Codeburst.io - Medium & Blog & [S55] \\ 
\hline
2019 & Measures to Microservices Security & OpensenseLabs & Blog & [S56] \\ 
\hline
2017 & Design Pattern: Microservice Authentication + Authorization & Keyhole Software & Blog & [S57] \\
\hline
\end{tabular}
}
\end{table*}

\renewcommand{\citenumfont}[1]{#1}
\renewcommand{\bibnumfmt}[1]{[#1]}

\section{Results} \label{sec:results}

This section present the results gathered from the systematic analysis of the final GL corpus. We start with an overview of the corpus composition (i.e., in terms of source types) and then proceed to the elaboration of the corresponding research questions.

\subsection{Corpus Overview}

\begin{figure}
\centering
 \includegraphics[width=\linewidth]{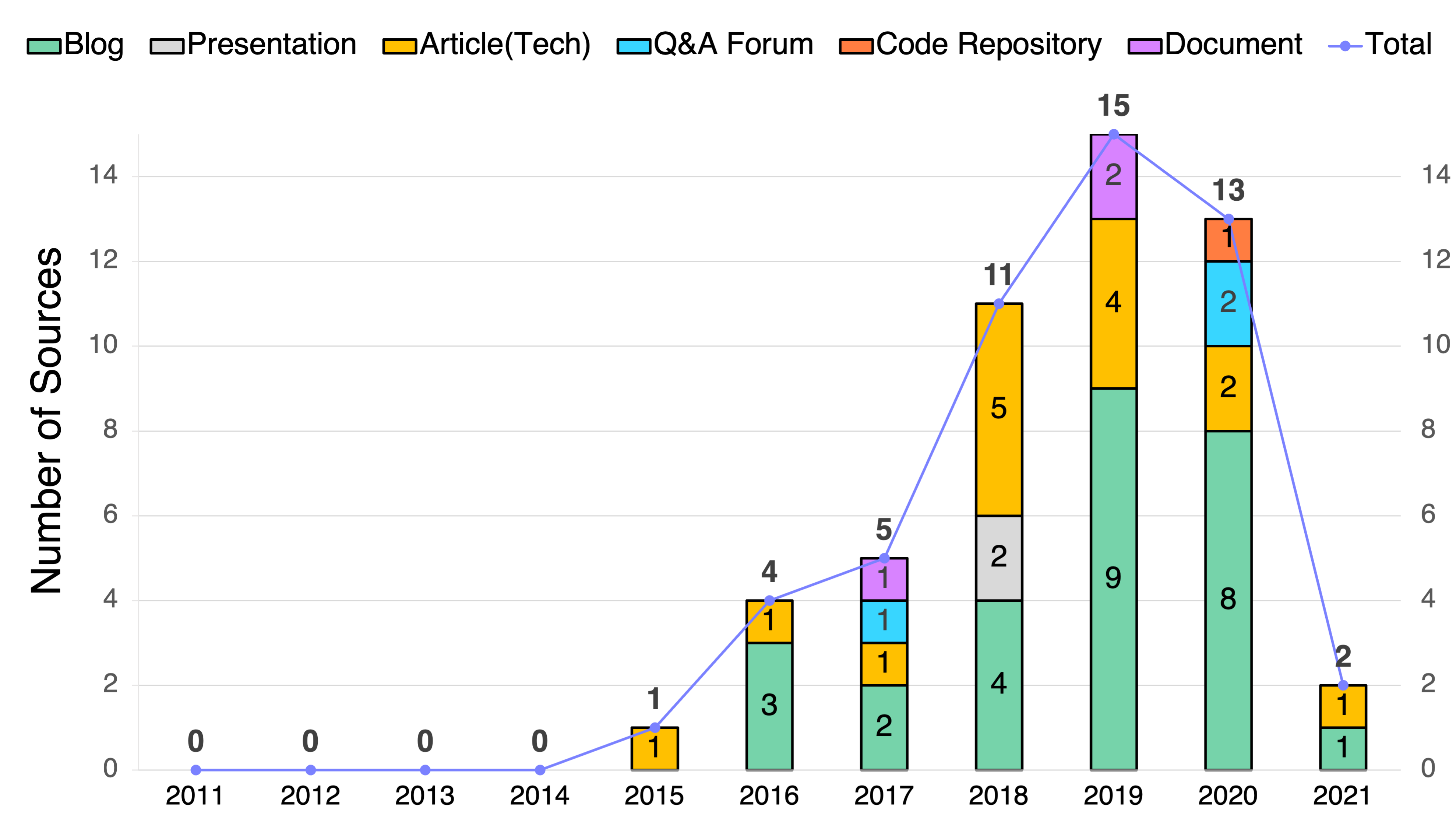}
\caption{Distribution of GL sources per type (2011 to 2021).}
\label{dist1}
\end{figure}

\autoref{dist1} shows the distribution of GL sources on microservice security from 2011 until 2021. Interestingly, the first references started to emerge in 2015, namely 4 years after the microservice concept was coined. This can be attributed to the novelty of this paradigm and also to the rise of new security threats. Most of the sources we analyzed correspond to \textit{blogs} (27 sources) and \textit{technical articles} (15 sources). The former are usually written in first person by the author, whereas the latter are expressed in a more journalistic style. \textit{Standards} and entries in \textit{Question \& Answer} (Q\&A) forums were also examined but with less quantity (3 sources each). Still, the presence of standards by major industrial security organizations such as OWASP and Cloud Security Alliance is quite valuable since it represents recent attempts to standardize microservice architectures. Likewise, Q\&A entries and \textit{code repositories} (1 source) are rich in terms of the technical challenges encountered by developers in the wild. Since 6 sources did not specify the year of publication, they are not shown in \autoref{dist1}. These include 4 \textit{blogs} and 2 \textit{presentations}. 

\subsection{Microservices Security Challenges (RQ1)}

To answer RQ1, we analyzed the security \textit{challenges} described within the selected GL sources, their \textit{causes} and \textit{consequences}. Furthermore, we examined the level of abstraction at which such challenges are encountered, as well as the security goals under threat.

\subsubsection{Security Challenges}

\begin{figure}
\centering
\includegraphics[width=\linewidth]{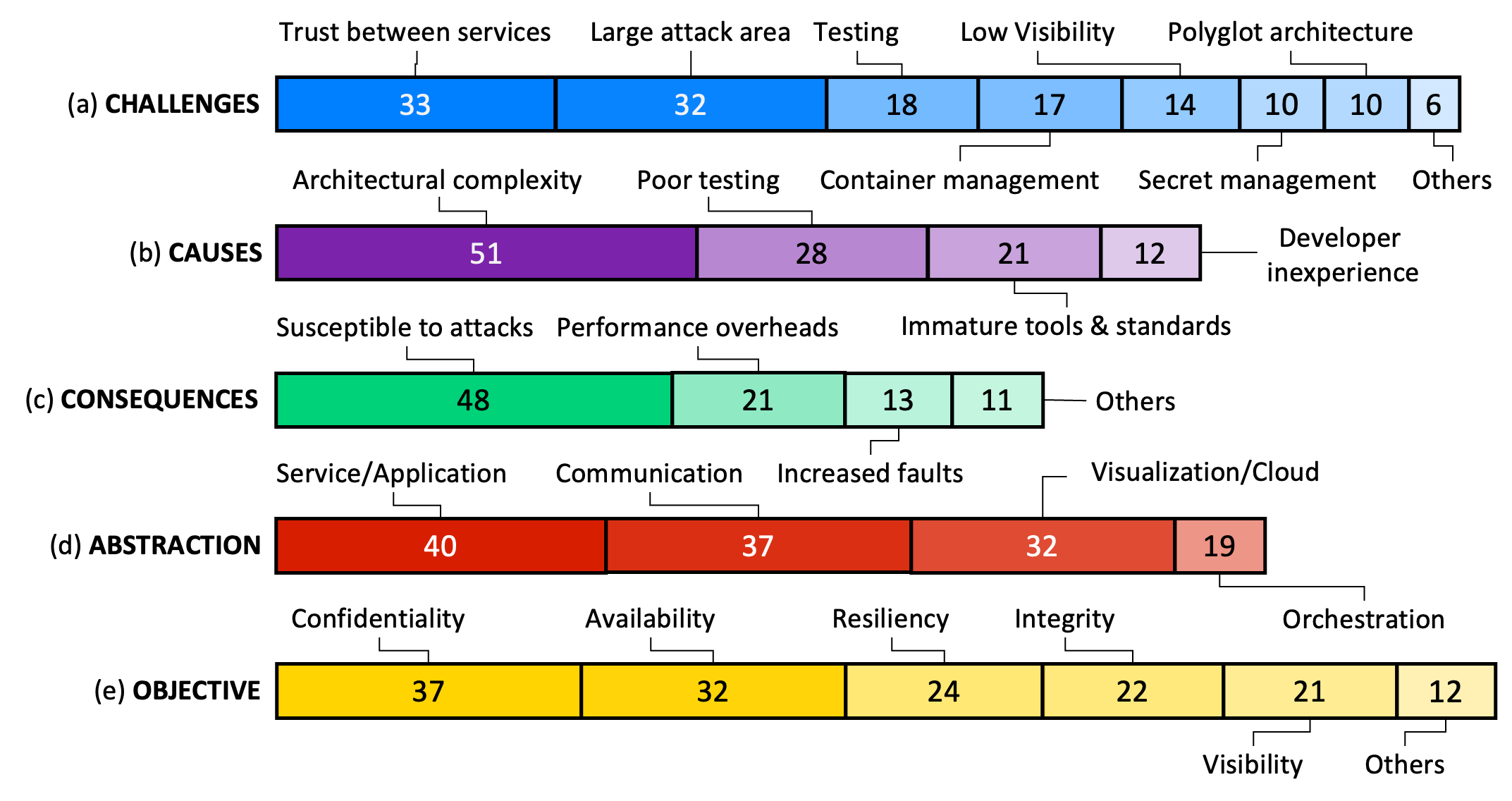}
\caption{Number of sources elaborating on security challenges (RQ1).}
\label{challenges}
\end{figure}

\autoref{challenges}(a) illustrates the most salient security challenges reported within the GL under analysis:


\begin{itemize}
    \item \textbf{Trust between services}: Performing a proper access control and identity management in a distributed fashion can be hard to achieve. Many GL sources express concerns about the limitations and vulnerabilities of current mechanisms. For instance, [S51] elaborates on the perils linked to Server-Side Request Forgery (SSRF) attacks, whereas [S58] and [S59] stress-out the difficulties of performing controls at the API level while handling multiple access requests. Hence, there is a call for shaping new authorization and authentication requirements in distributed architectures [S43].
    \item \textbf{Large attack area}: Several publications report challenges related to the multiplicity of attack points available in microservice architectures. This is due to their distributed and evolving nature (services are created dynamically) that makes their security hard to achieve through the traditional/monolithic methods (e.g., network segmentation) [S06, S07]. Loosely coupled services can also cause inter-service communication and data consistency issues [S08] making them prone to SQL injection attacks [S13].
    \item \textbf{Testing}: Unsurprisingly, GL sources reveal a lack of security testing in microservice architectures, mostly due to the speed of agile development and absence of automated test environments.
    \item \textbf{Container management}: A large number of vulnerabilities in containers have been reported since 2019 [S52]. Often, unsavvy developers use images available in public repositories that may not be verified nor properly configured [S5, S18]. This can lead to security breaches as attackers can leverage such images to gain access to application files and other containers [S34].
    \item \textbf{Low visibility}: Since microservices are not restricted to a single location, it is often hard to spot the exact source of sudden failures [S26]. Such a lack of visibility can lead to poor control over their infrastructure [S1, S51] and result in a slow recovery after a security breach [S6].
    \item \textbf{Secret management}: Several sources report problems when protecting secrets in microservice architectures. A lack of centralized storage, proper APIs, along with an inadequate management infrastructure are some of the issues reported within the GL [S9, S22]. Secrets are an easy target for lateral threat movement in the absence of default encryption and robust secret management [S51].
    \item \textbf{Polyglot architecture}: Deploying microservices using a multiplicity of programming languages introduces several layers of complexity as each of them have different life cycles and versions [S3, S9]. Particularly, it can be very challenging to implement a polyglot architecture without the right security expertise as every framework in the stack (along with their particular issues) must be treated differently [S30, S14].
    \item \textbf{Other} challenges include decentralized logging [S2, S8, S9, S24] and service mesh [S21, S38].
\end{itemize}

\subsubsection{Causes and Consequences}

\autoref{challenges}(b) highlights the various causes of the challenges mentioned above. On the one hand, the distributed complexity of microservice architectures is at the root of security challenges. Compound software solutions often entail bulky code that is prone to security attacks [S46]. On the other hand, agile microservice deployment tend to neglect security testing which translates into multiple flaws and vulnerabilities. Last but not least, developers' inexperience and oversight can be the cause of multiple security problems as they gain access to sensitive information inside microservices' code.

As shown in \autoref{challenges}(c), susceptibility to cyberattacks is one the most direct consequences of the microservice architectural style. Performance overheads in certificate configuration, management, and communication are also reported as a negative consequence in many GL sources, further emphasizing the importance of developing lightweight security solutions. Increased faults such as slow network connection, timeouts, or unavailability are also reported as consequences of security flaws in microservice architectures. Other drawbacks reported in GL sources are low traceability [S1, S2, S8] and system breakdown [S6]. 

\subsubsection{Security Goals and Concerns}
    
As shown in  \autoref{challenges}(d), many security concerns posited within the GL are related to issues encountered at the service/application layer. These include malicious code injection in the service discovery or service mesh infrastructure, faulty access control mechanisms, and sensitive information leakage. Concerns at the communication layer such as the transmission of confidential data are also discussed within the analyzed sources. Moreover, problems at the virtualization and orchestration layer were also identified. Surprisingly, there is no mention to security issues related to hardware in any of the analyzed sources. To a certain extent, this can be due to the fact that this layer is often less accessible to attackers \cite{yarygina2018overcoming}.

As shown in \autoref{challenges}(e) the difficulties mentioned above may hinder the achievement of several security goals in microservice architectures. Many sources reported issues when elaborating on aspects related to confidentiality, availability, and resiliency (24). This relates closely to the necessity of developing systems capable of dealing with security breaches in real time while keeping their responsiveness and reliability levels high. On the other hand,  integrity and visibility were also presented as desirable goals in microservices. The former refers to the ability of maintaining a system's internal consistency in the event of a security attack, whereas the latter is concerned with the possibility to closely monitor all security-related events within the system. Both are extremely important to promote efficient threat handling during an attack as well as to detect anomalies in the behaviour of microservices. Other microservice security goals emphasised by practitioners are the need for higher scalability [S5, S6, S7] and traceability [S2, S8, S9].
    
\subsection{Securing Microservices (RQ2)}

RQ2 elaborates on the solutions prescribed by practitioners to build secure microservice architectures. Particularly, RQ2a is concerned with best practices available within the analyzed GL, whereas RQ2b puts emphasis on technical solutions. We distinguish security best practices from technical solutions in the sense that the former primarily highlight the \textit{do's} and \textit{don'ts} of microservice security. Conversely, the latter are seen as recommendations which may acquire the form of (i) \textit{methods} (including standards, protocols, and specifications), (ii) \textit{tools}, and (iii) \textit{patterns}. 

\begin{figure}
\centering
 \includegraphics[width=\linewidth]{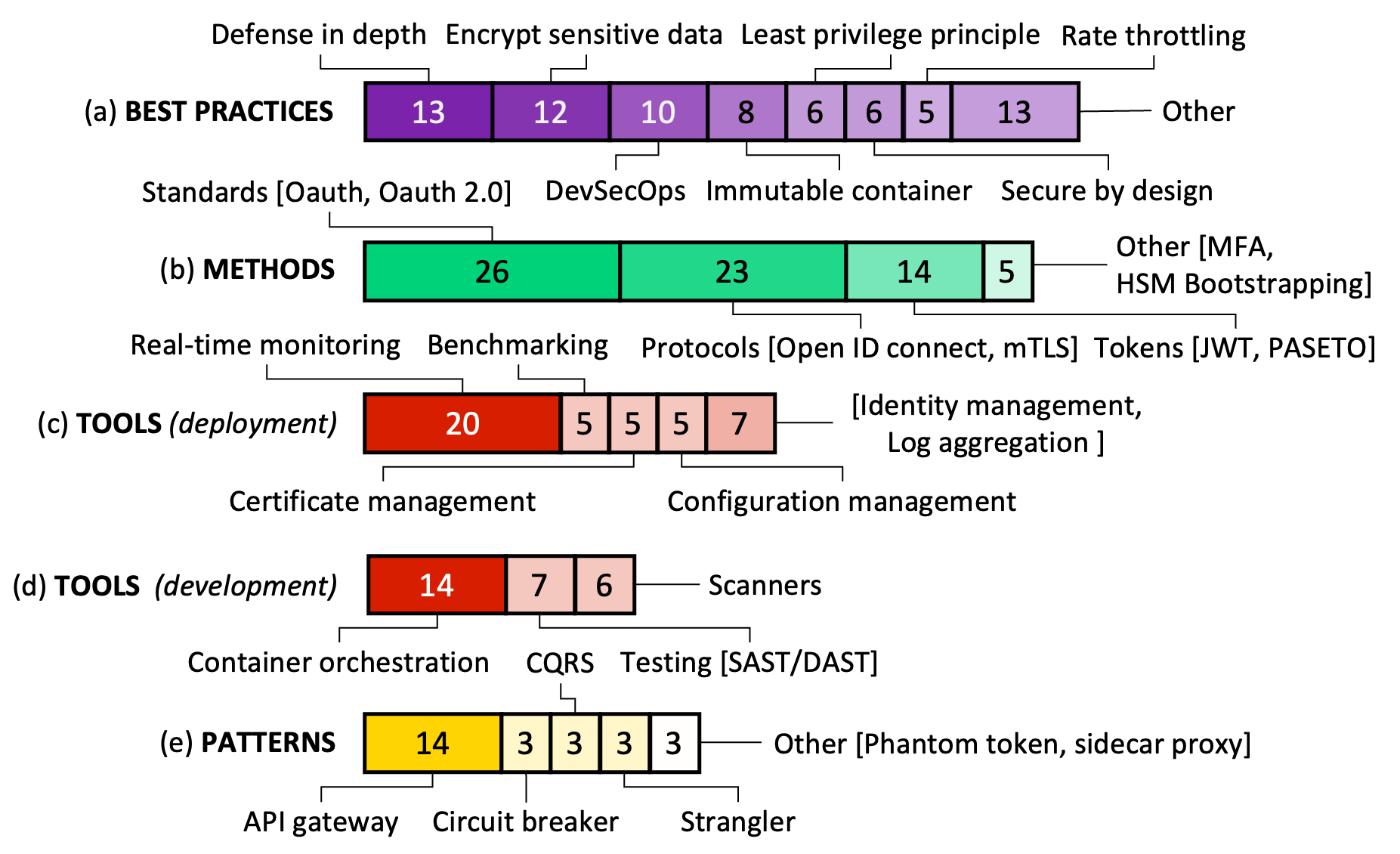}
\caption{Number of sources elaborating on security solutions (RQ2).}
\label{dist}
\end{figure}

\subsubsection{Best Practices (RQ2a)}

As shown in \autoref{dist}(a), these are the most salient security best practices we have identified:
\begin{itemize}
    \item \textbf{Defense in depth}: Sensitive resources should be safeguarded by multiple levels of security measures. This allows systems to stay protected in case a security mechanism fails (i.e., by activating backup/redundant defensive mechanisms) [S9, S17, S24].
    \item \textbf{Encrypt sensitive data}: Sensitive data resources should be encrypted so they remain confidential and only accessible to authorized people [S11].
    \item \textbf{DevSecOps}: 
    This approach seeks to incorporate security principles and standards throughout the whole software lifecycle while ensuring continuous integration and fast deployment. For this, it implements issue tracking to ensure timely identification of any defects along with continuous and automated security testing [S15, S24].
    
    \item \textbf{Immutable container}: Containers allow the development and execution of microservices across different computing environments. A good security practice consists of making such containers ``immutable'', which means that no updates can be performed after their deployment [S29, S31, S41]. In addition, storing data outside containers is highly recommended to ensure its access in case a container needs to be replaced.
    \item \textbf{Least privilege}: This refers to the way permissions are granted under a Role-Based Access Control (RBAC) schema [S14, S27, S49]. Particularly, it is important to ensure that each role has access to the least amount of resources. In other words, a user should only access the resources necessary to fulfill her tasks [S4, S47, S53].
    \item \textbf{Secure-by-design}: As mentioned in Section~\ref{intro}, security should not be an afterthought, but a central aspect throughout the whole microservice life cycle [S34, S46]. Secure-by-design works in tandem with DevSecOps to timely identify vulnerabilities in the development process [S24].
    \item \textbf{Rate throttling}: This approach consists of slowing down the response time of microservice applications if a deviation of its expected behaviour is detected [S14, S26].
    \item \textbf{Other} best practices include the use of reusable code [S12, S24, S37, S52], E2E testing [S35, S40], micro-segmentation [S26, S48], IP white-listing [S40, S41], and repave [S9].
\end{itemize}

\subsubsection{Technical Solutions (RQ2b)} There are several technical solutions discussed throughout the GL on microservice security. \autoref{dist}(b) shows the most salient \textbf{methods} for securing microservice architectures. Standardized \textit{authorization frameworks} such as OAuth and OAuth 2.0 are frequently recommended among practitioners [S14, S32, S35]. Likewise, \textit{protocols} like OpenID [S11, S22] and mTLS [S26, S41] are often discussed and presented as suitable solutions to authentication issues. The use of \textit{tokens} is also mentioned in various GL sources. For instance, the JSON Web Token (JWT) is regarded as a safe method for transferring claims between two parties [S19,  S40, S57]. However, JWTs can also introduce significant performance overhead, which is one of its major limitations. This point is better addressed by the Platform-Agnostic Security Token (PASETO), as it compresses JSON data into a single token that can be securely shared over the web [S14].

\autoref{dist}(c) depicts \textit{real-time monitoring} as one of the most explored tools for secure \textbf{microservice deployment} within the GL. Tools for \textit{benchmarking} are also regarded as effective means for ensuring a good baseline for system hardening [S9, S31]. Tools for  \textit{configuration-} and \textit{certificate-management} are further employed to support the centralized management of various roles, permissions, and certificates [S9, S22, S37, S43]. Likewise, \textit{identity management} tools are considered effective means for managing users, applications, and APIs in a centralized fashion while providing real-time monitoring [S6, S45]. \textit{Log aggregation} tools can also aid secure microservice deployment as they allow tracking all events from a single location as well as dispatching alerts in case of anomalous behaviour [S9, S44].

\begin{table}[h]
\fontsize{12}{12}\selectfont
\centering
\setlength{\tabcolsep}{5pt}
\def\arraystretch{1.7}
\resizebox{\linewidth}{!}{%
\begin{tabular}{| m{6cm} | m{10cm} |}
\hline
\rowcolor[HTML]{D5D3D2} 
\multicolumn{2}{|c|}{\textbf{Tools for managing microservice security}}
\tabularnewline \hline
Monitoring and Vulnerability Scanners &
  Prometheus, Clair, Statsd, Snyk**, Influxdb, Grafana, Twistlock, JFrog XRay** , BurpSuite**, OWASP Open RASP,Black Duck Hub \tabularnewline \hline
Configuration management Tool &
  Chef** ,Ansible, Puppet, Kubernetes, Docker \tabularnewline \hline
Image Scanning Tool &
  Clair, Anchore \tabularnewline \hline
Certificate Management Tool &
  Hashicorp Vault, Certbot, WildFly Elytron, Spring Vault \tabularnewline \hline
Container Orchestration Tool &
  Docker Bench, DC/OS, Istio, Azure Kubernetes Service, Vmware, Openshift** \tabularnewline \hline
Benchmarking Tool &
  CIS Docker Scanner \tabularnewline \hline
Testing Tool (SAST, DAST) &
  OWASP ZAP, HP Fortify**, Neuralegion NexDast** \tabularnewline \hline
Indentity Management Tool &
  AWS, Vmware, Octarine, Keycloak**  \tabularnewline \hline
Code Scanners &
  OWASP Source Code Analysis \tabularnewline \hline
Log Aggregation Tool &
  ELK Stack, Datadog**  \tabularnewline \hline
\multicolumn{2}{c}{ \underline{Note}: (**) Indicates proprietary software.}
\end{tabular}%
}
\caption{Tools proposed in the grey literature.}
\label{tools}
\end{table}

When it comes to \textbf{development tools}, \autoref{dist}(d) shows that \textit{container orchestration} is among the most discussed ones within the analyzed sources [S11, S33, S47]. Particularly, it is said to provide high levels of control specially in cases where regular software updates are necessary. Likewise, \textit{end-to-end testing} is presented as a security-enhancing tool during microservice development, being Static Application Security Testing (SAST) and Dynamic Application Security Testing (DAST) among the most popular ones [S1, S6, S14]. The use of scanners for detecting security vulnerabilities and flaws inside containers is also regarded as a useful instrument during development [S3, S12, S20, S31]. \autoref{tools} summarizes the different tools pointed by the GL sources.

According to the analyzed sources, \textbf{patterns} are often leveraged for building secure microservice solutions (\autoref{dist}(e)). \textit{API Gateway} for instance, is a applied in cases where data owned by multiple microservices must be fetched [S14, S35, S46]. Particularly, it encapsulates the system architecture into an API tailored for each individual thus acting as an authentication/authorization hub [S8, S21, S39]. \textit{Circuit breaker} on the other hand is a pattern that aims to prevent a chain reaction of security failures [S21]. For this, it sets a threshold which determines the maximum amount of failures a microservice should tolerate. The \textit{Command Query Responsibility Segregation (CQRS)} [S4] and \textit{strangler} patterns are also regarded as suitable for securing microservice architectures [S30]. The former consists of separating read from update operations in a data store, hence making it easier to ensure that only authorized entities write on such data. The latter on the other hand is used when migrating monolithic applications into a microservice architecture. Overall, it consists of gradually replacing specific functionalities with microservices until the legacy application is decommissioned~[S30].

\section{Overlooked Areas in Security Research} 
\label{sec:overlooked_areas}

Research in microservice security has also been documented extensively within white literature sources. In this section we contrast our results against the ones summarized in white literature reviews to identify \textit{overlooked areas} in microservice security research. That is, security aspects of microservice architectures that are mentioned by GL sources but have not been thoroughly investigated and documented in academic publications (and vice-versa). For this, we refer to the work of \citet{yarygina2018overcoming}, \citet{pereira2019security}, and \citet{hannousse2021securing} as these are well-aligned with the scope and aim of our work.

Some results of our study are aligned with findings reported by \citet{yarygina2018overcoming}. For instance, both works highlight the use of \textit{immutable containers} to safeguard microservices against security threats during deployment. Likewise,  both reported a set of technical solutions (e.g., mTLS, and JWT) along with security patterns and best practices (e.g., \textit{circuit breaker} and the \textit{principle of least privilege}). However, although \cite{yarygina2018overcoming} elaborates on the benefits of \textit{polyglot architectures} (i.e., a lower susceptibility to lateral attacks), it overlooks their security challenges (i.e., different life cycles, versions, and required expertise). Moreover, unlike \cite{yarygina2018overcoming}, our analysis reveals that container environments can introduce security threats (e.g., insecure images, poor authentication and authorization mechanisms), and summarizes orchestration management tools. However, \textit{N-version programming}, a method for dealing with the complexity of microservice architectures, is reported in \cite{yarygina2018overcoming} but not mentioned by any of the sources included in our GL corpus.

Likewise, the work of \citet{pereira2019security} detected 18 security mechanisms employed in microservice-based systems. These are focused primarily on authorization and authentication issues, along with means for preventing and detecting security attacks. Still, little attention is given to methods, patterns, and tools for reacting to and recovering from such attacks. Overall, the authors suggest that increasing developers' awareness can be considered as a suitable reactive measure. They also suggest auditing the tracing logs of the compromised system as a method for recovering from attacks. Conversely, our work suggests that the \textit{low visibility} of microservice architectures can make such an audit challenging (i.e., due to the decentralization of event logs). Furthermore, one of the GL sources we inspected says that recovering from an attack often entails destroying all dependency code to ensure that traces of viruses are removed from the system [S9]. Hence, there seem to be an offset between GL and white literature when it comes to attack recovery approaches in microservices.

Findings reported by \citet{hannousse2021securing} suggest that infrastructure-based attacks (i.e., those targeting monitors, discovery services, etc.) are not extensively addressed by academic sources probably due to their high complexity.  In line with this, our analysis reveals that \textit{rate throttling} can be used as a best practice against such attacks (most precisely against denial-of-service attacks). On the other hand, the authors of \cite{hannousse2021securing} also claim that research on internal attacks (i.e., those coming from insiders) also seem receiving little attention from white sources. In this regard, our analysis show that the \textit{principle of least privilege} can help preventing such attacks by granting minimum access permissions to the system's resources.

\vspace{2ex}
\begin{GrayBox}\small
\textbf{RESEARCH OUTLOOK.}
\begin{enumerate}[leftmargin=3ex]
\item The advantages of \textit{immutable containers} and \textit{polyglot architectures} are discussed in both GL and white sources. Still, related security challenges seem to be overlooked and thus require additional research efforts.\vspace{1ex}
\item White sources suggest auditing the tracing logs of compromised systems to recover from security attacks. However, this strategy is considered unfeasible by some GL sources due to the decentralization of event logs and the importance of deleting them to remove vulnerability traces.\vspace{1ex}
\item GL sources consider the \textit{principle of least privilege} and \textit{rate throttling} suitable practices for preventing internal and infrastructure-based attacks, respectively. They yet represent overlooked areas in security research demanding further attention and investigation efforts.
\end{enumerate}
\end{GrayBox}

\section{Limitations and Threats to Validity} \label{sec:threats-to-validity}

Based on the categories proposed by Wohlin et al. \cite{wohlin2012experimentation}, we identified the following limitations and threats to validity in our study:

\textit{Threat to internal validity:} The threat to internal validity refers to how well the evidence gathered justifies research questions.  As discussed in Section~\ref{preparation_phase}, we developed a strict inclusion and exclusion criteria to guide the selection of relevant references. To avoid ambiguity in the coding of the sources, a criteria catalogue was also maintained. Additionally, measures were taken to avoid external biases on the search process. 

\textit{Threat to external validity:} The treat to external validity encompasses all aspects that hinder the generalization of the study. The major threat to external validity is linked to the possibility of missing-out relevant terms in the search string. To mitigate this threat, a discussion on synonyms and applicable terms was conducted (Section~\ref{preparation_phase}).

\textit{Threat to construct validity:} The generalization of the experiment or methodology of the study is referred to as construct validity. This threat is reduced by stopping the search on saturation as described in Section~\ref{filtering_phase}. 

\textit{Threat to conclusion validity:} It refers to the ability to draw correct conclusions based on the information in the source. At each stage of GL analysis multiple feedback iterations were conducted to calibrate the results and conclusion of the work. 

\section{Conclusions and Future Work} \label{sec:conclusion}

This work presented a systematic study of GL sources and analyzed the security challenges practitioners come across when securing microservice architectures along with best practices and technical solutions. A wide variety of sources were identified for this study including, blogs, presentation, articles, Q\&A, code repository, documents and tools published between 2015 and 2021. We observed some areas of research and innovation in which industry and academia are aligned, specially when it comes to certain challenges and technical solutions (e.g., the use of immutable containers and security patterns). However, we also noticed some offset points and deviations regarding security threats in container environments and the use of N-version programming for handling the increasing complexity of microservice architectures. Moreover, practitioners and academic seem to have different opinions about the benefits and drawbacks of polyglot architectures, as well as on approaches for recovering from cyber-attacks. This calls for industry and academia to come together for shaping and fostering further research in securing microservice-based systems and elaborate on those points where different views meet. As a future work we plan to conduct an analysis of microservice security discussions in StackOverflow\footnote{\url{https://stackoverflow.com}} (i.e., by applying topic modelling) and compare them against the results obtained in our GL review.

\begin{acks}
This work was partly funded by the European Union's Horizon 2020 programme under grant agreement No. 952647 (AssureMOSS).
\end{acks}


\bibliographystyle{ACM-Reference-Format}
\bibliography{references}


\section*{GREY LITERATURE SOURCES}
\renewcommand{\bibsection}{}
\renewcommand{\bibnumfmt}[1]{[S#1]}

\end{document}